\documentclass[aps,prl,twocolumn,superscriptaddress,floatfix,amsmath,amssymb,longbibliography,10pt]{revtex4-1}

\usepackage[utf8]{inputenc}
\usepackage[T1]{fontenc}
\usepackage{txfonts}
\usepackage{tgtermes}
\usepackage[altg]{qtxmath}
\usepackage{braket}
\usepackage{graphicx}
\usepackage{amsmath}
\usepackage{amssymb}
\graphicspath{{figurem/}}
\usepackage{xcolor}
\usepackage[english]{babel}
\usepackage{microtype}
\usepackage{url}
\usepackage[normalem]{ulem}

\def\BBraket#1{\left\langle\!\left\langle#1\right\rangle\!\right\rangle}
{\catcode`\|=\active
  \xdef\BBraket{\protect\expandafter\noexpand\csname BBraket \endcsname}
  \expandafter\gdef\csname BBraket \endcsname#1{\begingroup
     \ifx\SavedDoubleVert\relax
       \let\SavedDoubleVert\|\let\|\BraDoubleVert
     \fi
     \mathcode`\|32768\let|\BraVert
     \left\langle\!\left\langle{#1}\right\rangle\!\right\rangle\endgroup}
}

\renewcommand*{\vec}[1]{\boldsymbol{#1}}

\makeatletter
\newcommand{\undersim}[1]{\mathrel{\mathpalette\@undersim{#1}}}
\newcommand{\@undersim}[2]{%
  \vcenter{%
    \ialign{%
      ##\cr
      $\m@th#1#2$\cr
      \noalign{\nointerlineskip\kern.2ex}
      $\m@th#1\sim$\cr
      \noalign{\kern-.4ex}
    }%
  }%
}

\makeatother

\begin{document}

\author{Q.~Z. Lv}
\email{qingzheng.lyu@mpi-hd.mpg.de }
\affiliation{Max-Planck-Institut f\"{u}r Kernphysik, Saupfercheckweg 1,  69117 Heidelberg, Germany }
\author{E.~Raicher}
\email{erez.raicher@mail.huji.ac.il }
\affiliation{Soreq Nuclear Research Center, 81800 Yavne, Israel }
\author{C.~H. Keitel}
\affiliation{Max-Planck-Institut f\"{u}r Kernphysik, Saupfercheckweg 1,  69117 Heidelberg, Germany }
\author{K.~Z. Hatsagortsyan}
\affiliation{Max-Planck-Institut f\"{u}r Kernphysik, Saupfercheckweg 1,  69117 Heidelberg, Germany }

\title{High-brilliance ultra-narrow-band x-rays via electron radiation in colliding laser pulses}

\date{\today}

\begin{abstract}

A setup of a unique x-ray source is put forward employing a relativistic electron beam interacting with two counter-propagating laser pulses in the nonlinear few-photon regime. In contrast to Compton scattering sources, the envisaged x-ray source exhibits an extremely narrow relative bandwidth of the order of $10^{-4}$, comparable with an x-ray free-electron laser (XFEL). The brilliance of the x-rays can be an order of magnitude higher than that of a state-of-the-art Compton source. By tuning the laser intensities and the electron energy, one can realize either a single peak or a comb-like x-ray source of around keV energy. The laser intensity and the electron energy in the suggested setup are rather moderate, rendering this scheme compact and table-top size, as opposed to XFEL and synchrotron infrastructures.

\end{abstract}

\date{\today}

\maketitle

\begin{figure}
  \begin{center}
  \includegraphics[width=0.475\textwidth]{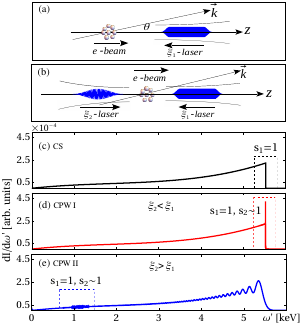}
  \caption{The setup of a relativistic electron beam colliding with a single laser pulse for CS (a), and with CPW (b). The emission spectra for CS (c), for Case I (d), and Case II (e). The electron energy is $\varepsilon=30\,m$ in all cases. $s_i$ denotes the absorbed photon numbers from the laser $\xi_i$ $(i=1,2)$, respectively.}
  \label{fig:demo}
  \end{center}
\end{figure}

Ever since the discovery by W. C. Röntgen in 1895,  powerful x-ray techniques have been developed
for determining the structure of matter at the atomic length scale \cite{Seeck_2014,Sanchez-Cano_2021}. Remarkable advancements have been achieved with the employment of synchrotron radiation \cite{Ternov_1995,Wille_1991,Nakazato_1989,Schlenvoigt_2008}, and the x-ray free-electron laser (XFEL) \cite{Chapman_2009,Pellegrini_2016,Bostedt_2016,XFEL,LCLS,Zhao_2017sxfel},  which dramatically increased the brightness of the source. Unfortunately, the large size and cost of these facilities,
limit their accessibility to a wide community. Alternative schemes rely on Thomson- and Compton-scattering (CS) \cite{Ting_1995,Schoenlein_1996,Sakai_2003,Gibson_2004}, and recently also on the radiation from laser-plasma interactions \cite{Rousse_2004,Corde_2013,Benedetti_2018,Sampath_2021}. The advancement of compact and powerful laser systems revived interest to these sources \cite{Hartemann_2010,Krafft_2010,Chang_2013,Chen_2013,Hartemann_2013,Sarri_2014,Khrennikov_2015,
Ghebregziabher_2013,Terzic_2014,Seipt_2015,Kharin_2018,Maroli_2018,Seipt_2019,Valialshchikov_2021,Debus_2010,Jochmann_2013}.
The CS source is based on a collision of a laser pulse with a relativistic electron beam [Fig.~\ref{fig:demo}(a)]. While not competing with brightness of large facilities, CS sources have several advantages, providing x-ray photons at a tunable energy in a broad spectral range, and being  relatively compact and affordable.

A compact brilliant x-ray light source with narrow bandwidth (BW) is an attractive tool, e.g., for x-ray imaging
in biology \cite{Carroll_2003}, x-ray nanoscale diagnostics in material science \cite{Wieland_2021,Li_2016,Hura_2009}, x-ray spectroscopy of highly charged ions \cite{Bernitt_2012,Kuhn_2020}. Recently a new field of x-ray quantum optics has been advanced aimed at the coherent control of atomic nuclei using shaped resonant x-rays \cite{Burvenich_2006,Rohlsberger_2012,Heeg_2017,Heeg_2021,Vagizov_2014,Chumakov_2018,Kuznetsova_2017}, which requires especially narrow  BW x-ray beams. Different schemes for narrowing the x-ray BW have been proposed involving temporal laser pulse chirping
\cite{Ghebregziabher_2013,Terzic_2014,Seipt_2015,Kharin_2018,Maroli_2018,Seipt_2019}, or temporally varying polarization \cite{Valialshchikov_2021} to compensate the nonlinear spectrum broadening. Alternatively, the x-ray photon yield at low BW can be enhanced using  a traveling-wave setup which allows an overlap of electron and laser beams longer than the Rayleigh length \cite{Debus_2010,Jochmann_2013}. However, all these approaches require the precise control of the pulse shape, phase or polarization, which is difficult in the high intensity domain.

In this Letter an alternative approach for narrow BW bright x-rays is put forward. Rather than modifying the laser pulse, an additional laser beam co-propagating with the electrons is introduced. Namely, the setup consists of a relativistic electron beam interacting with two counterpropagating waves (CPW) [Fig.~\ref{fig:demo}(b)]. The electron motion features two typical frequencies, separated by orders of magnitude because of the Doppler effect, $\omega_1=\omega_0(1+v_z)\approx 2\omega_0$ and $\omega_2=\omega_0(1-v_z)\approx\omega_0/2\gamma_*^2$ where $\omega_0$ is the laser frequency and $v_z$ the relativistic average velocity on axis and $\gamma_*$ the effective Lorentz factor \cite{Supp} (units $\hbar = c = 1$ are used throughout). Due to the nonlinearity of the relativistic dynamics the electron absorbs several photons in both frequencies in the considered regime when emitting an x-ray photon. As a result in the emission spectrum the Doppler-shifted high frequency $\omega_1$ peak is accompanied with satellites of $\omega_0$ separation. While the gross features of the spectrum (the spectral envelope) are determined by the counterpropagating laser beam, the  subtle features are governed by the second co-propagating laser beam. Accordingly, the BW of satellites scales with the smaller frequency $\omega_2$, allowing for bright ultranarrow BW x-rays.

The radiation has been calculated employing the Baier-Katkov semiclassical operator method \cite{Baier_b_1994}, applicable when the electron dynamics is quasiclassical in strong background fields. The radiation spectrum reads \cite{Lv_2021a,Lv_2021b}:
\begin{equation}
  \label{eq:bk_intensity}
  dI = \frac{\alpha}{(2 \pi)^2 \tau} \left[ -\frac{{\varepsilon'}^2 + {\varepsilon}^2 }{2  {\varepsilon'}^2} |\mathcal{T}_{\mu}|^2 +
          \frac{m^2 \omega'^2}{2 {\varepsilon'}^2 {\varepsilon}^2} |\mathcal{I}|^2 \right] d^3\vec{k'} \,,
\end{equation}
where $\mathcal{I} \equiv \int_{-\infty}^{\infty} e^{i \psi} \, \mathrm{d}t$, $\mathcal{T}_{\mu} \equiv \int_{-\infty}^{\infty} v_{\mu}(t) \, e^{i \psi} \, \mathrm{d}t$,   $\psi \equiv  \frac{\varepsilon }{\varepsilon'} k' \cdot x(t)$ is the emission phase, and $x_{\mu}(t)$, $v_\mu(t)$ , $k'_{\mu}=(\omega',\vec{k'})$ are the four-vectors of the electron coordinate, the velocity and the emitted photon momentum, respectively. $\tau$ is the pulse duration, $\varepsilon$ the electron energy with $\varepsilon' = \varepsilon - \omega'$.
In our setup  ultrarelativistic electrons  counterpropagate to the circularly polarized laser field with a vector-potential $A_1(x,t) = m \xi_1  [ \cos (k_1 \cdot x) e_x + \sin (k_1 \cdot x) e_y  ]$, where $\xi_1= e E_0 /(m \omega_0)$,
$E_0$ and $\omega_0$ are the laser field amplitude and frequency, respectively. $k_1 = (\omega_0,0,0,-\omega_0)$ is the laser four-wave vector with $\omega_0=1.55$~eV, and $e_x = (0,1,0,0), e_y = (0,0,1,0)$. $-e$ and $m$ are the electron charge and mass, respectively. The second laser field co-propagating with the electrons is also circularly polarized:  $A_2(x,t) = m \xi_2  [ \cos (k_2 \cdot x) e_x + \sin (k_2 \cdot x) e_y ]$, with
$k_2 = (\omega_0,0,0,\omega_0)$. The two lasers have the same frequency $\omega_0$ in the Lab-frame. We keep $\xi_1<1$, namely $\xi_1=0.1$, while choosing either $\xi_2<\xi_1$ (Case~I, $\xi_2=0.02$) or  $\xi_2>\xi_1$ (Case~II, $\xi_2=2$).
The quantum  strong-field parameter $\chi= e\sqrt{-( F^{\mu \nu}P_{\nu})^2}/m^3$, with the field tensor $F^{\mu \nu}$, and four-momentum $P_{\nu}$, is small $\chi\sim 10^{-5}$ for the chosen parameters, and
multiple photon emissions are negligible.

\begin{figure}
  \begin{center}
  \includegraphics[width=0.5\textwidth]{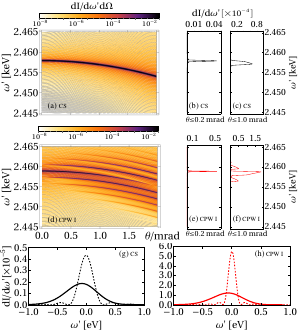}
   \caption{The angle-resolved spectra: (a) CS; (d) CPW Case~I. The angle integrated spectra: within $\theta \leq 0.2 $~mrad for (b) CS; (e) CPW Case~I and within $\theta \leq 1 $~mrad for (c) CS; (f) CPW Case~I. The spectra from a realistic electron beam (solid) and a monochromatic electron beam (dashed) : (g) for CS; (h) for CPW Case~I. The $x$-axis in panels (g,h) has been shifted by the peak energy in (b) and (e), respectively. The electron beam has a Gaussian distribution in both angle ($\Delta \theta=1$mrad) and energy ($0.01\%$ relative FWHM) domain with central energy $\varepsilon=20m$. The pulse length for $\xi_1$ is $16000$ cycles while the pulse length for $\xi_2$ is the same as the electron beam length, which is about $750$ cycles.}
     \label{fig:spec_angle1}
  \end{center}
\end{figure}

\begin{figure}
  \begin{center}
    \includegraphics[width=0.5\textwidth]{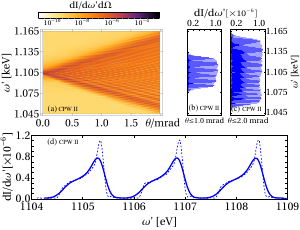}
  \caption{(a) The angle-resolved spectra for CPW Case~II. The angle integrated spectra for CPW Case~II: (b) within $\theta \leq 1 $~mrad; (c) within $\theta \leq 2 $~mrad; (d) The spectrum from a realistic electron beam for CPW Case~II (solid) and a monochromatic electron beam (dashed). The electron beam has the same distributions as in Fig.~\ref{fig:spec_angle1}, but with central energy $\varepsilon=30m$. The pulse length for $\xi_1$ is $8000$ cycles while the pulse length for $\xi_2$ is the same as the electron beam length, which is about $750$ cycles.}
  \label{fig:spec_angle2}
  \end{center}
\end{figure}

The radiation spectra are presented in Fig.~\ref{fig:demo}. The general features of all the spectra are similar. In the CS case, $\xi_2=0$ [Fig.~\ref{fig:demo}(c)], the spectrum has a sharp edge corresponding to absorption of a single photon (as $\xi_1<1$) from the laser field. The same edge is also dominant in spectra for cases I and II in CPW,
because $\chi$, determining the general spectral shape, is  dominated by the $\xi_1$-laser. However, the details of the spectra in CPW
reveal features stemming from the $\xi_2$-laser which are absent in CS.
For Case~I [Fig.~\ref{fig:demo}(d)] a single but ultra-narrow harmonic rises with a similar location but larger strength near the spectrum edge. For Case~II [Fig.~\ref{fig:demo}(e)] the entire spectrum becomes oscillatory and in most of the energy domain these oscillations are quite wide. However, the radiation emitted on axis, corresponding to $s_2 \sim 1$, exhibits a comb of sharp peaks, which is the main interest here and discussed below.

The angle-resolved spectra are presented in Fig.~\ref{fig:spec_angle1} for CS and CPW Case~I, and  Fig.~\ref{fig:spec_angle2} for CPW Case~II. While in the CS case a single harmonic appears in the given limited spectral range [Fig.~\ref{fig:spec_angle1}(a)], in CPW many satellites accompany the given harmonic  due to absorption of additional photons from the $\xi_2$ laser [Figs.~\ref{fig:spec_angle1}(d) and \ref{fig:spec_angle2}(a)]. Generally the radiation is distributed in a rather large angle and frequency region. Since $\xi_1<1$, the photon number for both CS and Case~I in CPW decreases monotonically in the $1/\gamma$-cone, while the BW increases with $\theta$ until it reaches a constant value \cite{Supp}. This makes the brilliance for both cases being the largest in the forward direction. Moreover, by integrating over the same angle range around $\theta = 0$, we can see that the radiation in Case~I is narrower in BW and more intense compared with CS [Fig.~\ref{fig:spec_angle1}(b,e)], which finally results in higher brilliance. After taking into account that the energy spreading and emittance of a realistic electron beam broaden the emission BW [Fig.~\ref{fig:spec_angle1}(g,h)], the advantage of Case I in brilliance compared to CS still remains at about one order of magnitude.


Furthermore,  a comb-like structure is produced in Case~II [Fig.~\ref{fig:spec_angle2}(b,c)]. While state-of-the-art techniques for a frequency comb can achieve the XUV domain \cite{Gohle_2005,Cingoz_2012,Cavaletto_2014,Porat_2018}, the covet is the hard x-ray regime. An attempt in this direction recently shown in \cite{Valialshchikov_2021} via CS with a polarization gating, demonstrated a relative BW of $10^{-2}$ and the spacing between peaks of the order of $\sim 100$ keV. In our CPW setup, however, the comb spacing is at an optical frequency, with more than one order of magnitude smaller relative BW [Fig.~\ref{fig:spec_angle2}(d)].

These results raise several questions: i) What determines the peak locations and the spacing between sequential peaks? ii) What determines the width of a single peak? iii) What is the role played by the angle window? and iv) How can we control the number of harmonics contained in the spectrum?
To address them, we turn to analytical estimations. The emission phase is a key variable
$\psi=\psi_{np}t-z_1  \sin \omega_1 t - z_2\sin \omega_2 t -z_3 \sin \Delta \omega_{12} t $, with $\psi_{np}=\varepsilon u (1-v_z \cos \theta)$, $z_1=(m \xi_1 u/\omega_1) \sin \theta$, $z_2=(m \xi_2 u/\omega_2) \sin \theta$, $z_3=2 \omega_0 m^2\xi_1\xi_2 u/(\varepsilon \Delta \omega_{12}^2)$, $\Delta \omega_{12}=\omega_1-\omega_2$, and $u=\omega'/(\varepsilon-\omega')$. The integral in Eq.~\eqref{eq:bk_intensity} over this phase yields multiplications of Bessel functions with arguments $z_1,z_2,z_3$ \cite{Lv_2021a,Lv_2021b}. From the phase follows the energy-momentum conservation, determining the emitted photon energy \cite{Supp}:
\begin{equation}
  \label{eq:bk_energy}
  \omega'= \frac{\omega^m_{s_1,s_2}}{1 + 2 \gamma_*^2 \left(1-\cos \theta \right) } \,,
\end{equation}
where $\omega^m_{s_1,s_2}= 2 \gamma_*^2 \left(  s_1 \omega_1 + s_2 \omega_2  \right)$ with $s_1, s_2$ being the numbers of photons absorbed from the first and second pulse, respectively, and $\gamma_* = \varepsilon/m_*$, with the effective mass  $m_*\equiv m\sqrt{1+\xi_1^2+\xi_2^2}$. The spacing between sequential $s_2$ harmonics according to Eq.~\eqref{eq:bk_energy} is $\Delta \omega'=2 \gamma_*^2 \omega_2\approx \omega_0$, because $\omega_1/\omega_2 \approx 4 \gamma_*^2$ and $\omega_1=2 \omega_0$. The rather broad emission frequency region in Figs.~\ref{fig:spec_angle1} and \ref{fig:spec_angle2} follows from Eq.~(\ref{eq:bk_energy}). Each line in Figs.~\ref{fig:spec_angle1}(d) and \ref{fig:spec_angle2}(a) represents the location of a single harmonic with respect to $\omega_2$. The spacing between adjacent lines is $\omega_0$.

To obtain a narrow BW radiation, one needs to apply the emission angle window ($\Delta \theta_w$). The width of the harmonic due to this finite angle window based on Eq.~\eqref{eq:bk_energy} is
\begin{equation}
  \label{eq:omega_theta}
  \delta \omega'_w=\omega^m_{s_1,s_2} \left( \gamma_*  \Delta\theta_w \right)^2 \,.
\end{equation}
In both cases, CS and CPW, the most beneficial for a small BW is the region near forward direction ($\theta=0$), when $\partial \omega'/\partial \theta=0$. Then, at applying a rather small angle window $\Delta \theta_w$, the BW is not $\Delta \theta_w$-dependent, but determined by the dynamical linewidth. The latter is smaller for CPW in comparison to CS.

The dynamical width of the harmonics is affected by two factors. The first stems from the characteristics of the electron dynamics. From the Bessel function features, the CPW harmonic width is estimated via $z_2 \approx 1$ \cite{Supp}. For the main peak
\begin{equation}
  \label{eq:omega_in}
  \delta \omega'_c = \frac{\omega_2}{8 } \left(\frac{m_*}{m\xi_2} \right)^2 \,.
\end{equation}
The latter BW scaling is surprising. In contrast to CS, the harmonics become narrower for increasing electron energy, confirmed with numerical calculations \cite{Supp}. For the considered parameters $\delta \omega'_c\ll \omega_0$, which results in an enhancement of satellite peaks over the spectral CS envelope. In fact, adding the second laser $\xi_2$ to the CS setup, new satellite peaks arise with a $\omega_0$ separation, and the smooth  energy distribution within a large BW for CS, $\delta \omega'_{CS}\sim \omega_0\gamma^2/\xi_1^2$ \cite{Supp}, roams into sharp spikes.

The second contribution to the harmonic width arises from the finite duration of the laser pulses. Note that the duration of the first pulse should be much longer, $N_1=n N_2$, where $n=\omega_1/\omega_2 $, and $N_{1,2}$ are the numbers of cycles experienced by a single electron during the interaction. Therefore, the photon uncertainty width is mostly determined by the second pulse
\begin{equation}
  \label{eq:omega_fin}
  \delta \omega'_{f} =  2 \gamma_*^2 \omega_2/N_2 = \omega_0/N_2 \,.
\end{equation}
Controlling the angle range $\Delta \theta_w$, one can tune $\delta \omega'_w$ mentioned above such as not to exceed the dynamical width: $ \delta \omega'_{in}={\rm max} \left( \delta \omega'_c, \delta \omega'_f \right)$ \cite{Supp}. Please note that the pulse length for $\xi_2$ and the electron beam in a realistic experimental setup should be similar such that all the electrons in the beam will contribute to the emitted x-ray pulse during the interaction \cite{Supp}.

In Case~II [Fig.~\ref{fig:spec_angle2}], the photon energy $\omega'$ is less sensitive to the emission angle than in Case~I because of a smaller velocity $v_z$ at the same energy (at different effective masses) and the number of harmonics is significantly larger than in Case~I because $m_*/m\xi_2 \approx \xi_1/\xi_2 \ll 1$. We can assess the effective number of harmonics $\Delta s_2$ for a given angle window $\Delta \theta_w$ by requiring $z_2/\Delta s_2 =\delta$, with a choice  $\delta \approx 0.8$, according to the Bessel function properties \cite{Supp}:
\begin{equation}
  \label{eq:harmonic_number}
   \Delta s_2 \approx \frac{1}{\delta} \left( \frac{ \Delta\theta_w}{  \theta_c} \right), \quad \theta_c=\frac{m_*}{8m \xi_2}\frac{1}{\gamma_*^3}
\end{equation}
where $\theta_c$ corresponds to the angle when $\delta \omega'_w (\theta_c)$ is equal to the characteristic width of the harmonics.
The integrated spectra over a different angle range in Fig.~\ref{fig:spec_angle2}(b,c) for Case~II reveals
a trade-off. On the one hand, the range of the comb can be extended by increasing the angle window.
On the other hand, it also induces a larger background in the gap between two sequential harmonics\cite{Supp}.
One should balance between the width of a single harmonic and the range of the total comb to have an optimized x-ray comb.

\begin{figure}
  \begin{center}
  \includegraphics[width=0.475\textwidth]{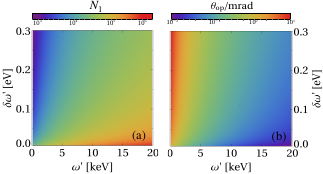}
  \caption{(a) the required pulse length $N_1$ of the $\xi_1$ laser for different emitted photon energies $\omega'$ with different harmonic peak widths $\delta \omega'$ in the spectrum. (b) the optimized angle range $\theta_{op}$ for different emitted photon energies $\omega'$ with different harmonic peak widths $\delta \omega'$ when we choose the pulse length in panel (a).}
  \label{fig:pulse_length}
  \end{center}
\end{figure}

The emittance and energy spreading of the electron beam will cause broadening of the radiation BW and suppress the brilliance. Is it possible with a realistic electron beam in the CPW setup to obtain brilliance better than in CS and generate the comb structure?  From Eq.~(\ref{eq:bk_energy}) we may estimate the contribution of energy $\Delta\gamma/\gamma$ and angular $\Delta \theta$ spread in the BW broadening as $\delta \omega'/\omega'\sim \max\{\Delta\gamma/\gamma, (\gamma \Delta \theta)^2\}$. Using electron beam parameters envisaged in \cite{deitrick_2018}: $\Delta\gamma/\gamma \sim 10^{-4}$ and $\Delta \theta \approx p_{\bot}/p_z \sim 10^{-3}$, we can expect the BW broadening up to $\delta \omega'/\omega'\sim 10^{-4}$.  We have tested our conclusion by numerical calculations of the spectra for an electron beam having a Gaussian distribution in both energy and angle domain with relative FWHM equal to $10^{-4}$ and $10^{-3}$, respectively [Figs.~\ref{fig:spec_angle1}(g,h),\ref{fig:spec_angle2}(d)]. Let us discuss the brilliance issue. The brilliance for both CPW and CS cases are the largest in the forward direction \cite{Supp}. In the case of angle window $\Delta \theta_w=0.2$~mrad in the forward direction [Fig.~\ref{fig:spec_angle1}(b,e)], the relative BW in CS for a single electron is $\Delta \omega'/\omega'\approx 10^{-4}$, while in CPW  Case~I it is 2 times smaller due to smallness of the dynamical BW. However, the number of photons in the line in the CPW case is about 7 times larger than in CS, which is due to more prominent forward emission in CPW. In the case of the electron beam with above parameters, the BW for CS is 1.4 times larger than for CPW, leaving however, the number of photons in the emission line equal to the case without spreading. Thus, we can have an order of magnitude larger brilliance in the CPW Case~I with respect to CS with the same laser and electron parameters. For instance using electron beam with 10pC charge and emittance of 0.1 mm$\cdot$mrad (with angle spreading $\Delta \theta=1$~mrad and transverse radius of $0.1$~mm) and $0.01\%$ energy spreading at electron energy $\varepsilon =10$ MeV \cite{deitrick_2018} the peak brilliance in forward direction for the CPW Case~I ${\cal B}\sim 5.2 \times 10^{18}$ ph/(s$\cdot$mrad$^2 \cdot$ mm$^2 \cdot$ 0.1\%BW), while in CS it is  ${\cal B}\sim 6.5 \times 10^{17}$ ph/(s$\cdot$mrad$^2 \cdot$ mm$^2 \cdot$ 0.1\%BW).

The comb-like structure is also preserved when using the electron beam with spreading parameters as above. Thus, Fig.~\ref{fig:spec_angle2}(d) shows that in the case of an angular window $\Delta \theta_w=2$ mrad the electron beam spreading induces a 2 times increase of BW. However, the comb structure, with numbers of peaks in the  comb $\Delta s_2=70$  is still preserved, characterized by the comb visibility, ratio of the BW to the peak separation,  being equal to $0.37$. The decrease of the angle window will improve the comb visibility, but will decrease the comb length.


Summarizing our approach, firstly, configuration I or II is chosen depending on the preference of the spectral shape: a single peak or a frequency comb. Then, the desired energy location and width are specified. The energy $\omega'$ determines the effective relativistic factor $\gamma_*$ via Eq.~\eqref{eq:bk_energy}. Then, from the chosen width $\delta \omega'$ and relation (\ref{eq:omega_theta}), the angle window $\theta_w$ follows. For Case~II, where $\delta \omega'_{f}$ dominates $\delta \omega'_{in}$, we also require $\delta \omega'_{f}=\delta \omega'$, from which the duration of the second pulse is evaluated. For a given laser  pulse energy and  the pulse duration  the field amplitude is derived, which determines the effective mass, and  from $\gamma_*$ one finds the electron energy [Fig.~\ref{fig:pulse_length}]. For a given BW, the number of cycles is determined according to Eq.~\eqref{eq:omega_fin}. However, increasing $\omega'$ requires a longer duration of the first pulse, due to increase
of $n=\omega_1/\omega_2 \approx 4 \gamma_*^2$. The optimal angle window $\theta_w$ according to Eq.~\eqref{eq:bk_energy} is narrower at higher $\omega'$.

In conclusion, the discussed CPW setup allows to generate an extremely narrow BW and collimated x-ray beam  in the range of several hundreds of eV to tens of keV, with brilliance by an order of magnitude exceeding that for
CS corresponding to the same laser and electron beam parameters. By tuning the intensity of the two laser pulses, one can produce either a single peak or a comb-like x-ray source.
This radiation source is attractive for several applications.
First, its narrow-band features allow for resonant excitation spectroscopy of highly charged ions \cite{Kuhn_2020,Bernitt_2012,Rudolph_2013,Lyu_2020,Nauta_2021}, without the use of a monochromator as opposed to synchrotron sources. Furthermore, by utilizing the comb, one can probe a much larger energy range in a single shot.
Second, the flux and BW of this source render it suitable to operate it as an XFEL seeder, thus replacing the complex and cumbersome self-seeding unit \cite{Nam_2021,Inoue_2019,Allaria_2013}, which is the only available seeding technique above energies of $100$ eV. Third, owing to the low angle spread it is favourable as a source for small angle scattering diagnostics \cite{Li_2016}. Fourth, the comb-like structure can be employed for extension of plasma diagnostics from the optical and UV range \cite{Dobosz_2005,Gibbon_1997,Theobald_1996} to the hard x-ray domain, e.g., for a temporally resolved measurement of the density profile of overdense plasmas. The proposed x-ray comb-structure  will enable probing much higher density plasma in a large density range in a single shot, with high accuracy due to the ultra-narrow BW of the single peak in the comb. Finally, replacing the laser fields with strong THz ones, the presented idea can be extended to a hard x-ray frequency comb for ultrahigh precision metrology.

Q.Z.L wishes to thank Zoltán Harman for helpful discussions regarding the applications of the x-ray frequency comb structure. We all thank J\"{o}rg Evers for useful comments on this Letter.

\bibliography{rad_xray}

\end{document}


\author{Q.~Z. Lv}
\email{qingzheng.lyu@mpi-hd.mpg.de }
\affiliation{Max-Planck-Institut f\"{u}r Kernphysik, Saupfercheckweg 1,  69117 Heidelberg, Germany }
\author{E.~Raicher}
\email{erez.raicher@mail.huji.ac.il }
\affiliation{Soreq Nuclear Research Center, 81800 Yavne, Israel }
\author{C.~H. Keitel}
\affiliation{Max-Planck-Institut f\"{u}r Kernphysik, Saupfercheckweg 1,  69117 Heidelberg, Germany }
\author{K.~Z. Hatsagortsyan}
\affiliation{Max-Planck-Institut f\"{u}r Kernphysik, Saupfercheckweg 1,  69117 Heidelberg, Germany }

\title{Supplemental Materials to the paper \\
\emph{"High-brilliance ultra-narrow-band x-rays via electron radiation in colliding laser pulses"}}

\maketitle

\section{The structure of the harmonics}
In this section, the details of the analytical estimations regarding the structure of harmonics, for example the harmonic width, the energy range of the x-ray comb and the optimized emitted anlge, are presented. All the estimations are based on the analytical formula of the emission spectrum as well as the energy conservation law, which is well documented in our previous works \cite{Lv_2021a,Lv_2021b}, and also verified by the numrical results in the main text.

The explicit formula for the radiation spectrum in general EM fields reads \cite{Baier_b_1994}:
\begin{equation}
  \label{eq:bk_intensity}
    dI = \frac{\alpha}{(2 \pi)^2 \tau} \left[ -\frac{{\varepsilon'}^2 + {\varepsilon}^2 }{2  {\varepsilon}'^2} |\mathcal{T}_{\mu}|^2 +
         \frac{m ^2 \omega^2}{2 {\varepsilon'}^2 {\varepsilon}^2} |\mathcal{I}|^2 \right] d^3\textbf{k}' \,,
\end{equation}
where $\mathcal{I} \equiv \int_{-\infty}^{\infty} e^{i \psi} \, \mathrm{d}t$ and $\mathcal{T}_{\mu} \equiv \int_{-\infty}^{\infty} v_{\mu}(t) \, e^{i \psi} \, \mathrm{d}t$ with $\psi \equiv  \frac{\varepsilon }{\varepsilon'}k' \cdot x(t)$ being the emission phase and $x_{\mu}$, $v_\mu$ , $k'_{\mu}=(\omega',\textbf{k}')$ are the four-vectors of the electron coordinate, the velocity and the photon momentum, respectively. $\tau$ is the pulse duration. Since the oscillations in the energy is small in amplitude \cite{Lv_2021a, Lv_2021b}, the average energy for the electron is $\varepsilon$, and $\varepsilon' = \varepsilon - \omega'$.

Let us first look at the phase $\psi$ of the emission, which is a crucial parameter in the formalism and determines the structure of the harmonics of the spectrum,
\begin{equation}
  \label{eq:bk_phase}
  \begin{split}
    \psi =& \frac{\varepsilon}{\varepsilon'} k' \cdot x(t) \\
         =& \psi_{np} t -  z_2 \sin(\omega_2 t) - z_1 \sin(\omega_1 t) - z_3 \sin(\Delta\omega_{12} t)    \,.
  \end{split}
\end{equation}
Introducing the definition of $u=\omega'/\varepsilon'$, $\omega_1=(1+v_z)\omega_0 \approx 2\omega_0$, $\omega_2=(1-v_z)\omega_0 \approx \omega_0/2\gamma_*^2$ and $\Delta\omega_{12}=\omega_2-\omega_1$ with $v_z$ being the average velocity on axis and $\gamma_* = \varepsilon/m_*$ where the effective mass $m_* \equiv m \sqrt{1+\xi_1^2+\xi_2^2}$, we may write
\begin{equation}
  \label{eq:bk_arguments}
  \begin{split}
    \psi_{np} \equiv& \varepsilon u(1 - v_z \cos \theta)   \,, \, \,
    z_1 \equiv \frac{ m u \xi_1 }{\omega_1 }\sin \theta\,,  \\
    z_2 \equiv& \frac{ m u \xi_2 }{\omega_2 }\sin \theta\,, \, \,
    z_3 \equiv \frac{2 \omega_0 m^2 u \xi_1 \xi_2 }{\varepsilon \Delta\omega_{12}^2} \cos \theta  \,.
  \end{split}
\end{equation}
With the analytical solution of the time integral in Eq.~\eqref{eq:bk_intensity}, one obtains
\begin{equation}
  \label{eq:bk_int_T}
  \mathcal{T}_{\mu} = 2 \pi \sum_{s_1} \sum_{s_1} \mathcal{M}_{\mu} (s_1,s_2,\omega',\cos \theta) \delta(\Omega_{s_1,s_1}) \,,
\end{equation}
where $s_1$ and $s_2$ denote the absorbed photon number from $\xi_1$ and $\xi_2$ pulse, respectively. The matrix elements can be expressed as the multiplication of Bessel functions with arguments $z_1$, $z_2$, $z_3$ \cite{Lv_2021b}. Furthermore, based on the argument of the $\delta$-function, the energy conservation can be written as
\begin{equation}
  \label{eq:bk_theta}
  \begin{split}
     \omega' =   & \frac{\varepsilon (s_1 \omega_1 + s_2 \omega_2)}{\varepsilon (1- v_z \cos \theta) +  (s_1 \omega_1 + s_2 \omega_2)} \\
         \approx & \frac{2 \gamma_*^2 (s_1 \omega_1 + s_2 \omega_2)}{1 + 2 \gamma_*^2 \left(1-\cos \theta \right) } \,.
  \end{split}
\end{equation}
The last step involves the approximation connected with the neglection of the recoil as $\chi \ll 1$. The emission is nearly on axis ($\theta \sim 0$) and can thus be confined to a small range $0<\theta<\theta_w$. The width of the harmonic due to this finite angle window is $\delta \omega'_w = \omega'(\theta=0)-\omega'(\theta= \theta_w)$. Employing Eq.~\eqref{eq:bk_theta} and Taylor expanding for $\theta \ll 1$ one obtains
\begin{equation}
  \label{eq:omega_theta}
  \delta \omega'_w=\omega_{s_1,s_2}^m \left( \gamma_*  \theta_w \right)^2 \,.
\end{equation}
where $\omega_{s_1,s_2}^m=2 \gamma_*^2 (s_1 \omega_1 + s_2 \omega_2)$.

\begin{figure}
  \begin{center}
  \includegraphics[width=0.48\textwidth]{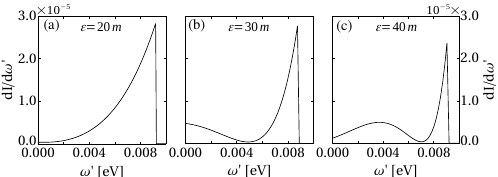}
  \caption{The spectra according to the analytical expressions of Eq.~\eqref{eq:bk_intensity} for $\xi_1=0.1$ and $\xi_2=2$ with different energies. In order to show the narrow width of the peaks, the x-axis has been shifted to the right by $494.928$eV for (a), by $1103.835$eV for (b) and by $1982.290$eV for (c). Here the spectra are calculated by the analytical formulas in Ref.~\cite{Lv_2021b}.}
  \label{fig:delta_omega_c}
  \end{center}
\end{figure}

As seen from the main text, the harmonics we are interested in are dertermined by the $\xi_2$-laser co-propagating with the electrons. The structure of the harmonics are therefore related with $z_2$ and the corresponding Bessel functions as well as the emission angle $\theta$. According to Eq.~\eqref{eq:bk_theta}, the corresponding sine function is approximately given by $\sin \theta \approx \sqrt{ 1 - 1/v_z^2 + 2 (s_1 \omega_1 + s_2 \omega_2 )/(\varepsilon u v_z^2)}$. Substituting this expression into Eq.~\eqref{eq:bk_arguments} one obtains
\begin{equation}
 \label{eq:est_z2}
    z_2 = \frac{\xi_2 m_* m}{\omega_2 \varepsilon} \sqrt{u(u_{s}-u)}\,,
\end{equation}
where $ u_{s} \equiv 2 \varepsilon  \left( s_1 \omega_1 + s_2. \omega_2 \right)/m_*^2$ . The main harmonics in Fig.~3 of the paper correspond to $s_1 = 1$ and $s_2 \sim 1$, which means $ u_{s} \approx 2 \varepsilon \omega_1/m_*^2$. Then, the width $\delta u$ can be estimated according to $z_2 \approx s_2$ in this regime and expressed as
\begin{equation}
 \label{eq:est_deltaU}
    \delta u = \frac{\omega_2^2 \varepsilon}{2\xi_2^2 \omega_1 m^2} \,.
\end{equation}
Therefore, according to $u=\omega'/\varepsilon'$, the intrinsic width of the peak $\delta \omega'_c$ is approximately
\begin{equation}
 \label{eq:est_deltaOm}
    \delta \omega'_c = \frac{\varepsilon^2 \omega_2^2}{2\xi_2^2 \omega_1 m^2+\varepsilon \omega_2^2} \approx \frac{\omega_2}{8} \left(\frac{m_*}{m \xi_2} \right)^2 \,.
\end{equation}
In the last step, we have neglected the photon recoil for simplicity. From this estimation, we can see that the width is decreasing with the increasing of the electron energy, seen in Fig.~\ref{fig:delta_omega_c}. According to Eq.~\eqref{eq:est_deltaOm}, the full width at half maximum of the peak is $0.00313$eV for panel (a), $0.00138$eV for panel (b) and $0.00076$eV for panel (c), which is approximately propotional to $1/\gamma_*^2$. This is in contrast with the width of the harmonic in the single plane wave case, where $\delta \omega' \sim \omega_0 \gamma^2/\xi^2 $ increases with the electron energy.

Meanwhile, we can also estimate the optimized emitted angle of the photons. Since the emission in our setup is almost on axis, the Bessel function argument in Eq.~\eqref{eq:bk_arguments} can thus be rewritten as $ z_2 \approx m \xi_2 u \theta/\omega_2 $. As the main peak appears when $u = u_s \approx 2 \varepsilon \omega_1/m_*^2$ and $z_2 \sim 1$, the optimized angle $\theta_{op}$ is
\begin{equation}
 \label{eq:est_theta_opt}
    \theta_{op} = \frac{ m_*^2}{8 \gamma \xi_2^2 \varepsilon^2} \,,
\end{equation}
which also decreases with the electron energy. This coincides with Fig.~4(b) in the main text. It is also worth to point out that this optimized angle is different from the scattering of an electron from a circularly polarized plane wave laser pulse with the optimized emitted angle around $\xi_1/\gamma$.


\begin{figure}
	\begin{center}
	\includegraphics[width=0.5\textwidth]{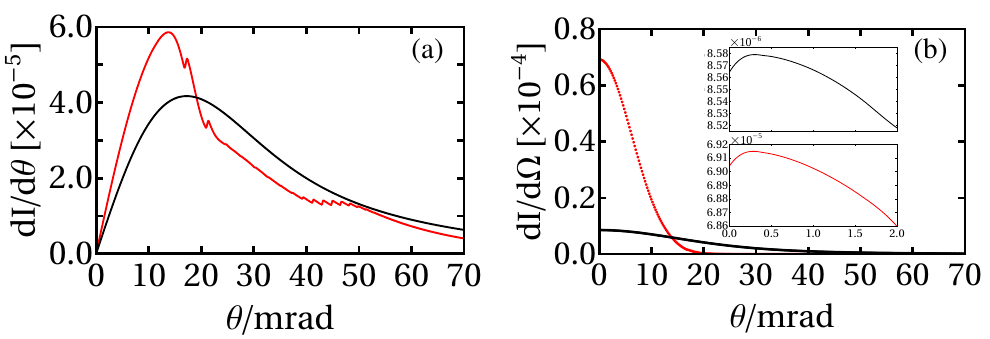}
	\caption{The angle-resolved radiation intensity for CS (black) and Case~I of CPW (red): (a) Intensity per emission angle $dI/d\theta$; (b) Intensity per solid angle $dI/d\Omega$. The insets in (b) show $dI/d\Omega$ for small $\theta$. All other parameters are the same as Fig.~2 in the paper.}
	\label{fig:figR1}
	\end{center}
\end{figure}

From $ z_2 \approx m \xi_2 u \theta/\omega_2 $, we can also see that $z_2$ tends to $1$ for the angle $\theta \sim 0$ as $\omega_2$ being rather small. This explains the almost on axis emission in the CPW setup. In the opposite, the harmonics in common CS, which triggered by $\xi_1$ only, is determined by $z_1$ and thus the harmonics center for larger angles according to Eq.~\eqref{eq:bk_arguments}.

Another interesting property of the spectrum is the energy range of the x-ray comb for the case of $\xi_2>\xi_1$. As explained in the paper, the peaks are equally distant with the laser basic frequency $\omega_0$, the estimation of this range is equivalent to estimating the number of peaks in the comb-like structure. The peaks are determined by Bessel functions with the argument of $z_2$, the same as for the intrinsic width. It is well known that a Bessel function tends to be zero when the argument is much smaller than the order of the function. Therefore, the peak appears only when $z_2/\Delta s_2 = \delta \sim 1$. With a given emitted angle window $0<\theta<\theta_w$ as before, $z_2$ can be approximated as $ 2 \xi_2 \theta_w \varepsilon (\Delta s_2 \omega_2 + \omega_1)/(\omega_2 m_*^2) $. Therefore, we can solve for $\Delta s_2$ based on $ z_2= \delta \Delta s_2$ and obtain
\begin{equation}
 \label{eq:est_s_R}
    \Delta s_2 = \frac{ 2 \xi_2 \varepsilon \omega_1 \theta_w}{\delta \omega_2 m_*^2 + 2 \xi_2 \varepsilon \omega_2 \theta_w}
            \approx \frac{1}{\delta} \left( \frac{ \theta_w}{  \theta_c} \right) \,.
\end{equation}
with $\theta_c=m_*/(8m \xi_2 \gamma_*^3)$ according to $\delta \omega'_w (\theta_c) = \delta \omega'_c$. From this formula we can see that the number of peaks is propotional to the angle spreading $\theta_w$ of the radiation.  If we choose the optimized angle in Eq.~\eqref{eq:est_theta_opt}, $\Delta s_2$ is propotional to $1/\xi_2$, which means that by lowering the strength of the $\xi_2$-laser in the CPW configuration, the range of the comb-like structure can be extended.


\begin{figure}
	\begin{center}
	\includegraphics[width=0.5\textwidth]{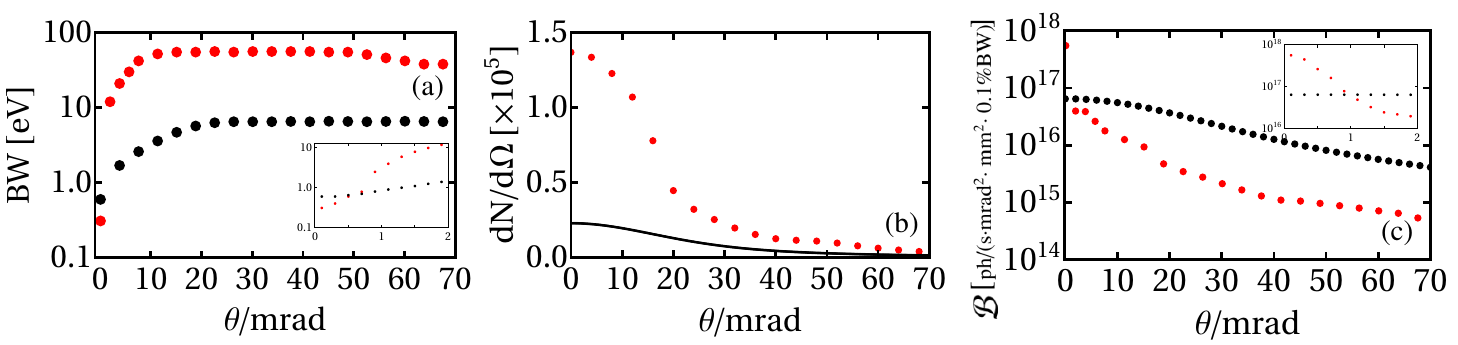}
	\caption{(a) The bandwidth (BW) of the spectrum as a function of angle $\theta$ for CS (black dots) and for Case I of CPW setup (red dots); (b) The photon number $dN/d\Omega$ as a function of angle $\theta$ for CS (black dots) and for Case I of CPW setup (red dots). The insets in (a) and (c) show the BW and brilliance for small $\theta$. Here the angle window is kept to be $0.2$mrad for all $\theta$ in the calculations. All the other parameters are the same as Fig.~2 in the paper.}
	\label{fig:figR8}
	\end{center}
\end{figure}

\section{Detailed comparison of common Compton scattering with the counter-propagating wave setup}
Let us clarify at which emission angle the brilliance is largest for CS with the laser and electron parameters used in the paper. From Fig.~1 in the main text, we can see that the general form of the spectra for all cases is the same as it is determined by $\xi_1$ laser and therefore the main radiation in Case I and II for the CPW setup is similar like in the CS case that distributes in a large angle range. We confined the emitted angle in a small range in order to achieve an extreme narrow BW. For the range of $\theta<0.2$~mrad in Case~I, the emitted energy is only $4.7 \times 10^{-5}$ of the total energy in all directions and this ratio becomes $1.2 \times 10^{-3}$ for $\theta<1.0$~mrad. This low relative portion is the price to pay in order to obtain monochromatic radiation and our scheme is more favorable for applications which require ultra-narrow band sources. 

In our calculations for normal CS and Case~I of CPW setup, the counter-propagating laser $\xi_1=0.1$, and therefore, the emission will be within the $1/\gamma$-cone. From Fig.~\ref{fig:figR1}, we can see that the intensity in a unit solid angle $dI/d\Omega$ for both CS and Case~I is largest in the forward direction and decreases monotonically with increasing $\theta$. The emission for Case~I is more concentrated in the forward direction because of the co-propagating laser pulse ($\xi_2$), but the total emission energy is similiar for both CS and Case~I as demonstrated in Fig.~1(c-e) of the paper.

The radiation BW for both CS and Case~I is shown in  Fig.~\ref{fig:figR8}. For small $\theta$ the BW for Case~I is smaller than that for CS. However, the BW for Case~I increases rapidly, when $\theta$ is increasing and then reaches a constant for large $\theta$. This is because  different harmonics of the $\xi_2$ laser overlap with each other for large $\theta$, and the BW in this case is determined only by the angle window, see the angle resolved spectrum in Fig.~2 of the paper. The angle dependence of the emitted photon number and brilliance are shown Fig.~\ref{fig:figR8}~(b) and (c). The brilliance decreases with $\theta$ for both cases. At large angles brilliance of CS is larger than for CPW Case I. However, in the forward direction Case~I shows larger brilliance with respect to CS, see inset in Fig.~\ref{fig:figR8}~(c). This advantage comes from the small BW and large photon number in the forward direction. Therefore, to show the advantage of our CPW scheme in brilliance with respect to CS, we will carry out a comparison at $\theta=0$.

In CS and the discussed CPW scheme the radiation is distributed in a rather large angle and frequency region, with a frequency-angle relationship according to the energy-momentum conservation [Eq.(2) of the paper]. To have a narrow bandwidth radiation, one needs to restrict the emission angle. In both cases, CS and CPW, the most beneficial for a small emission BW is the region near forward direction ($\theta=0$), when $\partial \omega'/\partial \theta=0$, see Fig.~2(a,d) of the paper. Then, when applying a rather small angle window $\Delta \theta_w$, the BW becomes not  dependent on the window, but is determined by the dynamical linewidth. At a given small angular window, the dynamical BW for CPW is significantly narrower than in CS, which is due to the properties of the spectral spikes in CPW. In the CPW setup new satellite peaks arise with $\omega_0$ separation, and the smooth energy distribution within a large BW for CS, roams into these sharp spikes. The  Figs.~2(b,e) of the paper show a comparison of the emission in CS with that of the CPW Case~I. Using the same angle window $\Delta \theta_w= 0.2$~mrad around $\theta=0$, the relative BW in CS for a single electron is $\Delta \omega'/\omega'\approx 10^{-4}$, while in the CPW  Case I it is 2 times smaller due to the smallness of the dynamical BW, see zoom in Figs.~2(g,h) of the paper (dashed lines are for a monochromatic electron beam). However, the number of photons in the line in the CPW case is about 7 times larger than in CS, which is due to more prominent forward emission in CPW, see Fig.~\ref{fig:figR1}~(b).

\begin{figure}
	\begin{center}
	\includegraphics[width=0.5\textwidth]{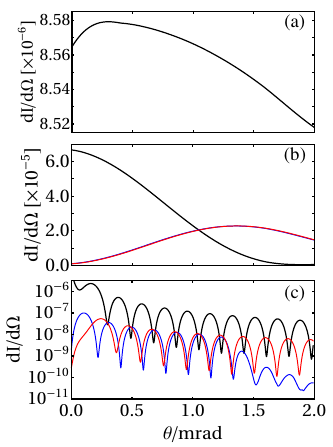}
	\caption{The angular distribution of radiation intensity: (a) CS; (b) CPW Case~I; (c)CPW  Case~II. In all the cases, $s_1=1$, $s_2=-1$ (blue curve), $s_2=0$ (black curve) and $s_2=1$ (red curve), for both CPW Case~I and Case~II. The  parameters are the same as in Fig.~2 in the paper.}
	\label{fig:figR3}
	\end{center}
\end{figure}

This can also be seen from the angular distributions of the radiation intensity for different harmonics in all cases as shown in Fig.~\ref{fig:figR3}. For CS [Panel (a)], we choose $s_1=1$, as at $\xi_1=0.1$ other harmonics are suppressed.  For CPW Case~I [Panel (b)] with $s_1=1$, the distribution is oscillating and the emitted energy for small angles is certainly larger than for common CS in Panel (a). By varying $s_2$, we can see that for different harmonics the emission is centered in different angle regions, which produces a single peak spectrum if we confine the angle range. For Case~II [Panel (c)], there are more oscillating in the same angle range for different harmonics and therefore it produces a comb structure in the frequency domain.

Due to these properties of the radiation, the brilliance for Case I in CPW is higher than that for CS. For a monochromatic electron beam the brilliance enhancement is approximately 14 $(7\times 2)$ times. In the paper we have considered also the case of a realistic electron beam with an energy spread and emittance.
We use the electron beam with the energy spread of the order of $\Delta\gamma/\gamma \sim 10^{-4}$ and an angular spread of the order of $\Delta \theta\sim 10^{-3}$, which is envisaged in \cite{Deitrick_2018}. In this case BW for CS is 1.4 times larger than for CPW, see  Figs.~2(g,h) of the  paper, leaving however, the number of photons in the emission line equal to the case without spreading. Thus, we can have an order of magnitude larger brilliance in the CPW Case~I with respect to CS with the same laser and electron parameters. For instance using an electron beam with 10pC charge and $0.1\%$ angle spreading (emittance of 0.1 mm$\cdot$mrad) and $0.01\%$ energy spreading at electron energy $\varepsilon =10$ MeV, the peak brilliance in forward direction for CPW Case~I is  ${\cal B}\sim 5.2 \times 10^{18}$ ph/(s$\cdot$mrad$^2 \cdot$ mm$^2 \cdot$ 0.1\%BW), while in the corresponding CS it is  ${\cal B}\sim 6.5 \times 10^{17}$ ph/(s$\cdot$mrad$^2 \cdot$ mm$^2 \cdot$ 0.1\%BW).

In order to obtain the largest brilliance for a realistic electron beam with a duration of $T_e$, each electron in the beam has to overlap with both of the laser pulses during interaction. The duration of the overlap of the electron beam with counter-propagating laser pulse is $\tau_1 \approx (T_1 - T_e)/2$ with $T_i=N_i T_L$ and the laser period $T_L$, which is also the total duration of the electron beam emission. During this time each electron should experience at least 5 cycles of the co-propagating laser beam: $(c-v)\tau_1> 5 c T_L$, i.e.
\begin{equation}
\label{eq:T1Te}
 T_1 - T_e > 20\gamma_*^2 T_L.
\end{equation}
All electrons in the beam should experience the co-propagating laser pulse, therefore, the electron beam should overlap with $\xi_2$ from the beginning of the interaction to the end and the length of the $\xi_2$-pulse should be longer than the electron beam, such that $c(T_2 - T_e)/(c-v)=\tau_1$. From the latter we derive the condition:
\begin{equation}
\label{eq:T2Te}
 T_2 - T_e>\frac{T_1}{4\gamma_*^2}.
\end{equation}
For the applied parameters in Case~I, $\gamma_*\approx 20$, $T_1=16000T_L$, and from Eq.~\eqref{eq:T1Te}, we have $T_e<8000 T_L= 20$ ps; $T_1/ (4\gamma_*^2)\sim 10T_L$, i.e., practically $T_2\gtrsim T_e$. We choose $T_2 \approx T_e=2$ ps and this choice is feasible as $\xi_2 \ll 1$. For Case~II where $\xi_2 \sim 1$, the pulse length or the transverse radius for the $\xi_2$ laser has to be smaller than the electron beam so that the laser can still be table-top size. This will decrease the brilliance but the comb structure in hard x-ray regime can be preserved. Hence, our scheme is feasible to operate with electron beams of a picosecond duration provided by usual electron accelerators, and using co-propagating long laser pulses of a picosecond duration. Note that the emitted x-ray pulse duration is $\tau_x=\tau_1 (c-v)/c+T_e \approx (T_1-T_e)/(4\gamma_*^2)+T_e\approx T_e$, which is taken into account in the estimation of brilliance.

\begin{figure}
  \begin{center}
  \includegraphics[width=0.48\textwidth]{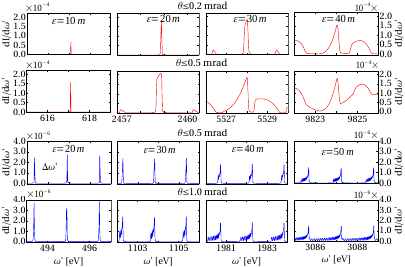}
  \caption{The spectral line's BW for different electron energies and different angle ranges: (upper panel, red curves) case I with $\xi_1=0.1$ and $\xi_2=0.02$; (lower panel, blue curves)  case II with $\xi_1=0.1$ and $\xi_2=2$. The pulse length for $\xi_1$ is $N_1=80000$.}
  \label{fig:energy_angle_various}
  \end{center}
\end{figure}

\begin{figure}[t]
  \begin{center}
  \includegraphics[width=0.48\textwidth]{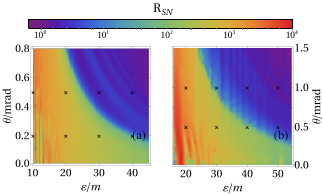}
  \caption{The signal-to-noise ratio $R_{SN}$ vs $\varepsilon$ and $\theta$: (a) for $\xi_1=0.1$ and $\xi_2=0.02$ and (b) for $\xi_1=0.1$ and $\xi_2=2$. The pulse length for $\xi_1$ is $N_1=80000$ for both cases. The oscillations in the plot stem from the background emission which induces oscillations in $I_{min}$, especially for case II. The crosses show the parameters considered in Fig.~\ref{fig:energy_angle_various}.}
  \label{fig:signal_noisy}
  \end{center}
\end{figure}

\section{The spectra for various energy and angle windows in the CPW setup}

As discussed in the main text, the BW of the spectrum in the CPW setup is determimed by three different factors. In Fig.~\ref{fig:energy_angle_various}, we show the interplay between different electron energies and emitted angle ranges. In case I, $m_*/(m \xi_2) \gg 1$, the dynamical BW  $\delta \omega'_{in}\approx \omega'_0$  is of the same order as the spacing $\Delta \omega'=\omega_0$ between peaks. Hence, enlarging $\delta \omega'_w$  results in gradually widening the peak. A different behaviour is seen for case II [lower panel], when $\delta \omega'_{in} \approx \delta \omega'_{f} \ll \omega_0$. Since $\delta \omega'_{f}$ does not depend on the energy and angle, the BW in case II originates from $\delta \omega'_w \sim \gamma^2$, which surpasses $\delta \omega'_{in}$ at $\varepsilon \gtrsim 30m $. For both cases in Fig.~\ref{fig:energy_angle_various} the relative BW of the harmonics in the spectra can be as small as $10^{-4}$ for both case I and case II, which is similar compared with that of XFEL.

In order to have a comprehensive picture of the impact of the electron energy and the emitted angle on the structure of the harmonics, we present the signal to noise ratio, $R_{SN}= I_{max}/I_{min}$, for the two cases in Fig.~\ref{fig:signal_noisy}(a) and (b). Here $I_{max}$ is the intensity at the main peak, while $I_{min}$ is the intensity at the middle point between the main peak and the adjacent one. By scanning over both $\varepsilon$ and $\theta$, we can see that the larger the electron energy, the smaller the angle should be such that the isolated peaks are sharp enough ($R_{SN}>100$). The sudden jump in $R_{SN}$ for both cases corresponds to the situation where the angle width reaches the middle point, $\delta \omega'_w=\omega_0/2$.

\section{The application of the comb-like structure in the x-ray regime}

The comb-like structure in the hard x-ray regime can be obtained by changing the intensity of the two lasers in the CPW setup. Compared with the comb in the EUV regime, the most energetic frequency comb available in the lab now, the comb-like structure in the x-ray regime is a completely new experimental tool and  can find applications in different fields. For example, it may be applied to probe ultra-dense plasma samples in a wide range of regimes such as inertial confinement fusion, laser-plasma interaction, as well as  laboratory astrophysics. The free electron density and the temperature of a dense plasma can be determined by measuring the absolute transmission of two harmonics from high-order harmonics generated in a gas jet, see Refs \cite{Dobosz_2005,Gibbon_1997,Theobald_1996}. The frequency comb described in this work enables probing plasma samples with much higher density as it lies in the x-ray domain. In addition, it dramatically improves the accuracy of the measurement since the single peak in the comb is ultra narrow and the relative spacing between two peaks is much smaller compared with radiation sources based on high-order harmonics generation. Moreover, the large frequency range of the comb allows for measuring a large density regime in a single shot. Secondly, the comb-structure can also be used in probing highly charged ion systems. Recent experiments, see Ref \cite{Rudolph_2013}, on determining the transitions in highly charged iron ions scanned over a large energy region employing a monochromatic x-ray laser pulse from XFEL facility. It is realized by applying a crystal monochromator. However, the frequency range of the monochromator is fixed by its crystal properties, and cannot be modified. On the contrary, the energy of the frequency comb proposed here may be freely tuned. Furthermore, by utilizing the comb, one can probe a much larger energy range in a single shot. On the other hand, because the electron energy and the laser intensity needed are rather moderate, this setup is of table-top size, relatively affordable and anticipated easy to operate.

\bibliography{rad_xray_supp}